\documentclass[aps,prd,11pt,onecolumn,groupedaddress,showpacs,nofootinbib,amssymb
]{revtex4}
\usepackage[dvips]{graphicx,color}
\usepackage{amssymb}
\usepackage{amsmath}
\usepackage{graphicx}
\usepackage{amsfonts}
\usepackage{bm}
\newcommand{\be}{\begin{equation}}
\newcommand{\ee}{\end{equation}}
\newcommand{\bea}{\begin{eqnarray}}
\newcommand{\eea}{\end{eqnarray}}
\newcommand{\beaa}{\begin{eqnarray*}}
\newcommand{\eeaa}{\end{eqnarray*}}

\allowdisplaybreaks[4]

\begin{document}

\title{Realization of Holographic Entanglement Temperature for a Nearly-AdS boundary}

\author{D. Momeni}

\address{Eurasian International Center for Theoretical Physics and Department of
General\\ \& Theoretical Physics, Eurasian National University, Astana 010008, Kazakhstan\\
momeni-d@enu.kz}
\author{M. Raza\footnote{corresponding author}}

\address{Department of Mathematics, COMSATS Institute of Information Technology, Sahiwal 57000, Pakistan\\
mreza06@gmail.com}

\author{H. Gholizade}

\address{Department of Physics, Tampere University of Technology P.O.Box 692, FI-33101 Tampere, Finland\\
hosein.gholizade@gmail.com}

\author{R. Myrzakulov}

\address{Eurasian International Center for Theoretical Physics and Department of
General\\ \& Theoretical Physics, Eurasian National University, Astana 010008, Kazakhstan\\
rmyrzakulov@gmail.com}

\tolerance=5000
\begin{abstract}
Computing the holographic entanglement entropy proposed by Ryu-Takayanagi shows that thermal energy near boundary region in $AdS_3$
gain maximum of the  temperature. The absolute maxima of
temperature  is  $T^{Max}_{E}= \frac{4G_3 \epsilon_{\infty}}{l}$. By
simple physical investigations  it has  become possible to predict a
phase transition of first order at critical temperature $T_c\leq
T_{E}$. As they predict a tail or root towards which the AdS space ultimately tend, the boundary is
considered thermalized. The Phase transitions of this form have received  striking
theoretical and experimental verifications so far.
\end{abstract}


\maketitle
\section{Introduction}	

In a pioneering work\cite{Bekenstein:1980jp}, it was shown that the entropy to energy
ratio of any closed system never exceeds the following bound (in which $G=c=\bar{h}=1$):
\begin{eqnarray}
\frac{S}{E}\leq 2\pi R.
\end{eqnarray}
In this formula, $R$ denotes the size (effective size) of the
system. From the statistical point of view, the existence of this
upper bound is transformed to the value of $\beta_0=T_{0}^{-1}$ such
that $Z(\beta_0)=0$, where $Z(\beta)$ is the partition function of
the system. All our errors in explaining the origin of entropy arise
from our obstinacy in believing that gravitational entropy is
entirely similar to thermodynamical one, which can be realized
statistically. The only possible reply is already given by
treating the origin of gravitational entropy as area
\cite{area1},\cite{area2}. It is evident that the area law is more
or less corrected with quantum effects, having its origin in
backreactions from the quantum process of  matter with which the
entropy is usually changed. Although  area of origin and its
appearance like the "boundary" of the spacetime is wholly modern,
yet its constitutional origin is analogous to that of the weakly
coupled gravitational models in bulk and the strongly coupled
quantum systems on boundary. Thus one may wonder if there is an
analogous relation for general quantum systems which are far from
equilibrium and a corresponding gravitational dual.  Most probably
this story had its origin in a particular gauge-gravity dual theory
as to the meaning of the entropy mistletoe. The origin of the area
law for entropy as a special type is wholly known  using the
celebrated AdS/CFT correspondence \cite{Maldacena,Gubser,Witten}.
The pioneering work of Ryu-Takayanagi showed vast development so far
about the computation of an entanglement entropy of a quantum system
 holographically \cite{RT1,RT2} (in section (2), we'll review the idea
and methodology). Their fame title, however, is their
pioneering work in the application of the AdS/CFT  to applied
research. During these years we became convinced that the success of
the Ryu-Takayanagi machinery in a field of condensed matter was
not to be reckoned with the tale of doubtful conversions. It is
opening up and starting its work on new grounds \cite{Peng:2014ira}-\cite{Bhattacharyya:2013gra}. \\
 Our aim in this paper is the production of a holographic entanglement
temperature $\frac{1}{T_E}=\frac{\Delta S}{\Delta E}$ , by means of
the undulations of holographic entanglement  entropy $\Delta S $ on
a near AdS region caused by energy $\Delta E$, of an infinitesimal
layer. \par

\section{ Holographic Entanglement Entropy (HEE)}

We suppose there are some "entangled" quantum systems in $N-1$ dimensions on boundary
which would considered to be divided  into two parts $A$ and $B$. The word "
total density " $\rho_{tot}$, which is the name used in statistical
mechanics for the  density operator of $A\cup B$, had originally a
general meaning, and may have required qualifications when applied
to this particular subsystems. But it has now become a specific
label, and the prefix "total" should be dropped. We now define the
reduced partial density matrix on a subsystem $A$ as
$\rho_A=\mbox{Tr}_B\rho_{tot}$ \cite{CaCa},\cite{Calabrese:2005zw}.
It is convenient here to define the entanglement entropy (EE) of
entangled subsystem $A$ which characterize entanglement between
$A$ and $B$ of the quantum system $A\cup B$ by $S_A=-\mbox{Tr}\rho_A\log
\rho_A$. This definition satisfy the subadditivity of the entropy
$S(A\cup B)\geq S_A+S_B$. It was worse than the AdS/CFT after
Ryu-Takayanagi discovered that $S_A$ of a quantum system in boundary
can be computed in the bulk gravity dual via the simple minimal area
functional (see figure 
\begin{eqnarray}
S_A=\frac{\mbox{Area}(\gamma_A)}{4G_N}\, \label{HEE}.
\end{eqnarray} here
$\gamma_A$ is a minimal (in the case of non static bulks it must be replaced by the maximal) area surface which has the same boundary
as $A$, i.e. $\partial\gamma_A=\partial A$ \cite{RT1,RT2}.

\par
\section{ Near $AdS_3$ boundary  calculations  }
The following metric and coordinates $(t,\rho,\phi)$ have been
officially adopted by global representation of $AdS_3$:
\begin{eqnarray}
ds^2=l^2\big(-\cosh^2\rho dt^2+d\rho^2+\sinh^2\rho d\phi^2)\label{g}.
\end{eqnarray}
The global form of the $AdS_3$ metric given by (\ref{g}) is the
official standard of full covered $AdS_3$ and is in use into the
holographic set-up of the Ryu-Takayanagi  algorithm. The AdS
boundary was located in the $\rho=\infty$. Instead, we replace it
with the finite radius $\rho=\rho_0$ to avoid divergence. Such
cutoff $\rho_0$, gives a high degree of simplicity to the
computation, but leave it as an open question whether it has the
exactitude of the expression of the HEE, or it is only the
approximation. This cutoff is valuable as enabling us to fix
approximately the expression of the HEE, which must have occurred
somewhere about $\rho_0\gg1$. The metric in this limit is,
approximately, $ds^2\simeq l^2 e^{-2\rho_0}(-dt^2+d\phi^2)$. It
provides the topology $\mathcal{R}\times S^1$ for surface defined by
$\rho=\rho_0\sim-\ln(\frac{2\pi l}{L})$. We're assuming the entangled quantum system "living" near to this cylinder with total length $L$ and characteristic length $l$. The physics are also described by $CFT_2$ as well as $AdS_3$ bulk geometry.\\
We parameterize the minimal bulk surface $\gamma_A$ as
$\gamma_A=\{t=t_0,\phi\in(0,\frac{2\pi l}{L}),0\leq \rho\leq
\rho_0\}$. The first step is to calculate the wrapped surface
(curve) $\rho(\phi)$ which represents the path between the  points
of the boundary separated by $\phi=0,\phi=\frac{2\pi l}{L}$. Taking
the Euler-Lagrange equation of motion  for the function
$\phi(\rho)$, we can calculate the following values for the extremal
functions corresponding to certain assigned values of integration
constants $\big(a,\phi_0\big)$:
\begin{eqnarray}
\coth(\rho)=a\cos\Big((a^2-1)(\phi-\phi_0)\Big)
,\ \ \rho(0)=\rho(\frac{2\pi l}{L})=\rho_0\label{rho-eq}.
\end{eqnarray}
Then by solving this equation, regarding the two elements $(a,\phi_0)$ as unknown quantities, the values of the parameters may be computed:
\begin{eqnarray}
\text{class I}:\ \ a^2=1-\frac{nL}{l},n\in[1,\infty),\ \ \phi_0=\frac{l}{nL}\arccos\Big(\frac{\coth(\rho_0)}{1-\frac{nL}{l}}\Big),\\
\text{class II}:\ \ a^2=1+\frac{2n\pi}{\frac{2\pi l}{L}-2\phi_0},n\in[1,\infty),\ \ \sqrt{1+\frac{2n\pi}{\frac{2\pi l}{L}-2\phi_0}}\cos\Big[\frac{2n\pi\phi_0}{\frac{2\pi l}{L}-2\phi_0}\Big]=\rho_0.\\
\end{eqnarray}
By substituting the "\text{class I}" parameters of  $(a,\phi_0)$ in (\ref{rho-eq}) we obtain:
\begin{eqnarray}
\text{class I}:\ \ \coth(\rho)=\sqrt{1-\frac{nL}{l}}\cos
\Big[\frac{nL}{l}\phi-\arccos\Big(\frac{\coth(\rho_0)}{1-\frac{nL}{l}}\Big)\Big], \ \ n\in[1,\infty)\label{rho-sol}.
\end{eqnarray}
Hence, solution (\ref{rho-sol}) is a disconnected surface. A graph
of this functions is plotted in (\ref{rho}). The difficulties
confronted in the way of solving "\text{class II}"  are very severe,
and up to the present time we are not agreed to the result. One can
solicit the possibility of solving "\text{class II}"  by pure
numerical methods. We'll use the "\text{class I}"  for further
studies.

\section{Near $AdS_3$ boundary  calculations}

The following metric and coordinates $(t,\rho,\phi)$ have been
officially adopted by global representation of $AdS_3$:
\begin{eqnarray}
&&ds^2=l^2\big(-\cosh^2\rho dt^2+d\rho^2+\sinh^2\rho d\phi^2)\label{g}.
\end{eqnarray}
The global form of the $AdS_3$ metric given by (\ref{g}) is the
official standard of full covered $AdS_3$ and is in use into the
holographic set-up of the Ryu-Takayanagi  algorithm. The AdS
boundary was located in the $\rho=\infty$. Instead, we replace it
with the finite radius $\rho=\rho_0$ to avoid divergence. Such
cutoff $\rho_0$, gives a high degree of simplicity to the
computation, but leave it as an open question whether it has the
exactitude of the expression of the HEE, or it is only the
approximation. This cutoff is valuable as enabling us to fix
approximately the expression of the HEE, which must have occurred
somewhere about $\rho_0\gg1$. The metric in this limit is,
approximately, $ds^2\simeq l^2 e^{-2\rho_0}(-dt^2+d\phi^2)$. It
provides the topology $\mathcal{R}\times S^1$ for surface defined by
$\rho=\rho_0\sim-\ln(\frac{2\pi l}{L})$. We're assuming the entangled quantum system "living" near to this cylinder with total length $L$ and characteristic length $l$. The physics are also described by $CFT_2$ as well as $AdS_3$ bulk geometry.\\
We parameterize the minimal bulk surface $\gamma_A$ as
$\gamma_A=\{t=t_0,\phi\in(0,\frac{2\pi l}{L}),0\leq \rho\leq
\rho_0\}$. The first step is to calculate the wrapped surface
(curve) $\rho(\phi)$ which represents the path between the  points
of the boundary separated by $\phi=0,\phi=\frac{2\pi l}{L}$. Taking
the Euler-Lagrange equation of motion  for the function
$\phi(\rho)$, we can calculate the following values for the extremal
functions corresponding to certain assigned values of integration
constants $\big(a,\phi_0\big)$:
\begin{eqnarray}
&&\coth(\rho)=a\cos\Big((a^2-1)(\phi-\phi_0)\Big)
,\ \ \rho(0)=\rho(\frac{2\pi l}{L})=\rho_0\label{rho-eq}.
\end{eqnarray}
Then by solving this equation, regarding the two elements $(a,\phi_0)$ as unknown quantities, the values of the parameters may be computed:
\begin{eqnarray}
&&\text{class I}:\ \ a^2=1-\frac{nL}{l},n\in[1,\infty),\ \ \phi_0=\frac{l}{nL}\arccos\Big(\frac{\coth(\rho_0)}{1-\frac{nL}{l}}\Big)\\&&
\text{class II}:\ \ a^2=1+\frac{2n\pi}{\frac{2\pi l}{L}-2\phi_0},n\in[1,\infty),\ \ \sqrt{1+\frac{2n\pi}{\frac{2\pi l}{L}-2\phi_0}}\cos\Big[\frac{2n\pi\phi_0}{\frac{2\pi l}{L}-2\phi_0}\Big]=\rho_0.
\end{eqnarray}
By substituing the "\text{class I}" parameters of  $(a,\phi_0)$ in (\ref{rho-eq}) we obtain:
\begin{eqnarray}
&&\text{class I}:\ \ \coth(\rho)=\sqrt{1-\frac{nL}{l}}\cos
\Big[\frac{nL}{l}\phi-\arccos\Big(\frac{\coth(\rho_0)}{1-\frac{nL}{l}}\Big)\Big], \ \ n\in[1,\infty)\label{rho-sol}.
\end{eqnarray}
Hence, solution (\ref{rho-sol}) is a disconnected surface. A graph
of this functions is plotted in (\ref{rho}). The difficulties
confronted in the way of solving "\text{class II}"  are very severe,
and up to the present time we are not agreed to the result. One can
solicit the possibility of solving "\text{class II}"  by pure
numerical methods. We'll use the "\text{class I}"  for further
studies.


\par
We have calculated the
changes of HEE, for the domain denoted by
$\mathcal{D}=\{\rho\in(\rho_0-\Delta\rho,\rho_0),\phi(\rho_0)=\phi(\rho_0-\Delta)\}$,
and change in the radial coordinate  by $\Delta\rho$, then the
approximate changes of HEE and energy are respectively given by the
following:


%

\begin{eqnarray}
&&\Delta S\simeq\frac{\Delta \rho}{4G_3}\Big(1+2c^2e^{-2\rho_0}+\mathcal{O}(e^{-4\rho_0})\Big),\\&&
\Delta E\simeq \frac{\epsilon_{\infty} \Delta \rho}{l}\Big(1-2e^{-2\rho_0}+\mathcal{O}(e^{-4\rho_0})\Big),
\end{eqnarray}
here $|c|^2=|-\frac{l}{nL}|$.
We used the Tolman law to compute the energy measured by a local
observer:
\begin{eqnarray}
&&E=\frac{\epsilon_{\infty}}{\sqrt{-g_{tt}}}\label{Energy}.
\end{eqnarray}
We may define this energy by the terms of  red-shifted  energy
$\epsilon_{\infty}$ at AdS infinity to the local observer.
The local energy of $AdS_3$ is always decreasing, and
never increasing; in contrast to the strictly decreasing.


From these expressions, by the help of near AdS approximation
$\rho\simeq \rho_{0}\gg1$, we can calculate the ratio $\frac{\Delta
S}{\Delta E}$ for any domain $\mathcal{D}$:
\begin{eqnarray}
&&\frac{\Delta S}{\Delta E}\simeq\frac{l}{4G_3 \epsilon_{\infty}}\Big(1+2(c^2+1)e^{-2\rho_0}\Big)\geq \frac{l}{4G_3 \epsilon_{\infty}}\label{ratio}.
\end{eqnarray}
Therefore there exist a minima in the value of the entanglement
entropy to energy ratio, depending on the $l,G_3,\epsilon_{\infty}$.
We can calculate the effective entanglement temperature of formation
$\mathcal{D}$ from its CFT boundary for any substance dissolved in a
given region $\mathcal{D}$, from the knowledge about $\frac{\Delta
S}{\Delta E}$, by means of an application of the well-known
thermodynamical process:
\begin{eqnarray}
&&T_{E}\leq \frac{4G_3 \epsilon_{\infty}}{l}\label{T}.
\end{eqnarray}
Which also gives us a "Universal" upper bound of temperature of the
near boundary region when the holographic entanglement entropy of
its boundary  is known. In this work, the holographic temperature
(\ref{T}), according to infinitesimal change of the holographic
entanglement entropy, add energy for a near AdS boundary domain
$\mathcal{D}$, we sought to demonstrate the upper bound of
temperature. It is an attempt once more to demonstrate that entropy
of the holographic system on boundary can be increased as a function
of energy. In statistical mechanics, the entropy of a closed system
$S(E)$ always increasing or remaining constant, and never
decreasing. Indeed, we  contrast this property of $S(E)$ with
strictly increasing. The same patterns serve to demonstrate the
upper bound condition of the quantum entanglement temperature of the
system in dual CFT.
 It's a lot different than a bound proposed in \cite{Bhattacharya:2012mi}, although, the
range of the computed holographic temperature for the  near-AdS
region  is both vast and subjective, but our bulk and its near
boundary geometry is so different from \cite{Bhattacharya:2012mi}
that trying to define the temperature for the excited states in the
CFT. But in reality it is not different from the "Universality"
relation $T_E\propto l^{-1}$. That is what we expect to be able to
do for more general geometries in excited states, because it is
theoretically possible in a systematic different way. Now we can
have something completely different: \par \emph{Universal upper
bound on the entanglement temperature is proposed as an attempt to
obtain a lower bound on the entanglement entropy to energy ratio of
a near-AdS region of space when the bound is  demographically
different from the one which was proposed  for the excited states}.
\par As an illustrative of this upper bound temperature, it may be
explained that any entangled quantum system such as investigated
here, will go well upon phase transitions, while the free energy $F$
set well. We compute the free energy $F$ of $AdS_3$ for the layer in
the limit of $e^{\rho_0}\gg1$ using the saddle point approximation
$e^{-\beta F}=\int{\rho(E) e^{-\beta E}dE }$ , we obtain:
\begin{eqnarray}
&&F(\beta,l)=-\frac{2G_3}{l}\beta-\frac{l}{2G_3\beta}-\frac{1}{2\beta}\ln\Big(\frac{l\pi}{4G_3}\Big)+\frac{\ln\beta}{\beta}-\frac{1}{\beta}\ln\Big(Erfi\Big[\sqrt{\frac{G_3}{l}}(a+\frac{l}{2G_3})\Big]\Big)
\end{eqnarray}
here $\rho(E)$ is the density of distinct energy states in the entangled system,  $\beta=\frac{dS}{dE}|_{E=E^*}$, $E=a \beta^{-1}, S_A(E)=S_0-\frac{l}{2G_3}\ln\Big(\frac{lE}{2\epsilon_{\infty}}\Big)$ \cite{RT1,RT2} and $Erfi[x]$ is the  imaginary error function is defined by
     $Erfi[x]=- i erf( ix)=\frac{2}{\sqrt{\pi}}\int_{0}^x e^{t^2}dt$.\par
According to the reference \cite{Jaeger}, we can classify the phase
transitions based on free energy. At quantum based phase transition
point (QPTP) first or second order derivative of free energy
respected to thermodynamics variables diverges \cite{Blundell}. In
our problem the first derivative of free energy respected to
temperature gives the entropy and second derivative  is proportional
to heat capacity. The entropy and heat capacity are:
\begin{eqnarray}
S=-\frac{\partial F}{\partial T}|_{V}\\
C_v= T \frac{\partial S}{\partial T}|_{V}.
\end{eqnarray}
There is a critical temperature $T_c$, that entropy becomes zero at this temperature. Below the $T_c$ the entropy has negative value and by approaching to zero temperature we may expect that the equation of state of system will change:
\begin{equation}
T_c=\frac{2 \sqrt{\frac{G_3}{l}}}{ \sqrt{W\left(\pi  e^{\frac{l}{G_3}+2} \text{Erfi}\left(a \sqrt{\frac{G_3}{l}}+\frac{ \sqrt{l}}{2\sqrt{G_3}}\right)^2\right)}}\label{TC}
\end{equation}
$W$ is  $W$-Lambert function and is defined by the solutions of the
equation $W(x)e^{W(x)}=x,\ \ x\in\mathcal{R}$ , $T_c\leq T_{E}$.\par

\emph{The Phase transitions of this form have received a striking
theoretical \cite{Faraggi:2007fu} and experimental verifications in
so far as they predict a tail or root towards which the AdS space
ultimately tend when the boundary is considered to be thermalized. }
The second part of the statement of our paper- the reality of the
prediction by phase transitions - has been frequently called in
question, chiefly on the ground that, in order to predict a strongly
coupled system on CFT with any chance of success, one should have
the command of certain thermodynamical facts which are known until
today, and then merely approximately, and only employed with that
object in the this paper. The question, however, is whether HEE
could predict the phase of the entangled system in CFT with any
chance of success - much less whether it could state beforehand at
what temperatures the phase would be visible, as some have
erroneously supposed, and which of estimates would have been quite
impossible for us to do. It is different, however, with physical
properties, density, etc, at present we have no fixed rules which
enable us to predict quantitatively the differences in physical
properties corresponding to a given difference in entangled system
in CFT and its near boundary dual system, the only general rule
being that those differences are not large. We cannot predict with
any exactness about the characters of a single phase individual
here, but if we consider mixing of large number of phases in this
nearly boundary layer, we can predict with considerable accuracy the
percentage of phases which will have the mean character proper to
their thermalization, or will differ from that mean character within
any assigned limits. Others, however, are random, that is to say,
the sequence of phases is repeated at irregular intervals, and it is
thus impossible to predict when the maximum and minimum HEE will
occur. We have not heard anything predicting even the possibility of
these random phases  before they came upon near-AdS  region.
\par
\section{Discussion}

It may be useful to summarize here the main results which have been
gained in this paper. If we define the holographic entanglement
entropy $S$ for a near boundary region of $AdS_3$, as the
thermodynamic entropy in which a small change in the radius of the
space charged with positive $\Delta S$ would tend to move, we can
observe an upper bound on the holographic temperature $T_{E}$ in a
simple form $T_{E}\leq \frac{4G_3 \epsilon_{\infty}}{l}$ by saying
that, if we have any minimal surface described in any manner in the
near boundary region, the excess of the amount of  the entropy flow
which leave the surface
$\frac{dS}{d\rho}\Delta_{\rho}|_{\rho=\rho_0}$ over the lack of
energy which enter it $\Delta E=E(\rho_0)-E(\rho_0-\Delta\rho)\sim
\Delta\rho \frac{dE}{d\rho}$ is equal to $\frac{l}{4G_3
\epsilon_{\infty}}$-times the algebraic sum of all the positive
infinitesimal terms included within the minimal surface. The study
of such cases suggests that the statement in terms of critical
phenomena of the transitions between the mixed phases of entangled
systems in the dual CFT may be only a critical one at $T_c\leq T_E$
in the bulk, which, though it may describe the effect of the actual
transitions between them sufficiently for some practical purposes,
is not to be regarded as representing them purely. In the limits
assigned to this paper it is impossible to enter further into the
nature of the phase transitions, but an attempt may be made to
summarize the holographic results so far as they bear upon the CFT
picture, which has again been revived in some branches, as to the
fundamental descriptions  of  strongly coupled systems.

\section*{Acknowledgments}

The work of M. Raza is supported by Higher Education Commission Pakistan by the project No.\emph{20-2497/NRPU/R\&D/HEC//12/7018}.


\end{document}